\documentstyle[12pt,epsf]{article}
\begin{document}

\title{Effects of Random Density Fluctuations on Matter-Enhanced
Neutrino Flavor Transitions in Supernovae and Implications for
Supernova Dynamics and Nucleosynthesis}
\bigskip
\author{F. N. Loreti, Y.-Z. Qian,\\
 G. M. Fuller\footnotemark[1], and
A. B. Balantekin\footnotemark[2]\\
Institute for Nuclear Theory,\\
 University of Washington,
     Box 351550, Seattle, WA 98195\\}
\renewcommand{\thefootnote}{\fnsymbol{footnote}}
\footnotetext[1]{Permanent
address: Department of Physics, University
of California, San Diego, La Jolla, CA 92093-0319.}
\footnotetext[2]{Permanent address: Department of
Physics, University of Wisconsin, Madison, WI 53706.}

\date{}
\maketitle

\begin{abstract}
We calculate the effects of random density fluctuations on
two-neutrino flavor transformations (${\nu}_{\tau(\mu)}
\rightleftharpoons {\nu}_e$) in the post-core-bounce supernova
environment. In particular, we follow numerically the flavor evolution
of neutrino states propagating through a stochastic field of density
fluctuations. We examine the approach to neutrino flavor
depolarization, and study the effects of this phenomenon in both the
early shock reheating epoch and the later $r$-process nucleosynthesis
epoch.  Our results suggest that significant fluctuation-induced
neutrino flavor depolarization effects occur in these environments
only when the zero-order (without density fluctautions) evolution of
the neutrino states includes adiabatic propagation through
resonances \\
(mass level crossings).  In the shock reheating
epoch, depolarization
effects from fluctuations with amplitudes larger than 0.05\% of the
local matter density can cause an increase in the heating rate of the
material behind the shock compared to the case with no neutrino flavor
transformation - but this corresponds to a significant decrease in
this quantity relative to the case with adiabatic neutrino
transformation.  If $r$-process nucleosynthesis is to occur during the
late stages of supernova evolution, then the requirement of
neutron-rich conditions excludes a region of neutrino mass-squared
difference and vacuum mixing angle ($\delta m^2,\ \sin^22\theta$)
parameter space for neutrino flavor transformation. We find that in
the presence of stochastic fluctuations, this excluded region is not
significantly altered even for random fluctuations with an amplitude
of 1\% of the local matter density.
\end{abstract}

\vfill
\eject

\section{INTRODUCTION}

In this paper we investigate matter-enhanced neutrino flavor
transformations, the MSW (Mikheyev-Smirnov-Wolfenstein) effect [1,2],
for the case of a matter density consisting of a smooth component and
a fluctuating part modeled by Gaussian colored noise.  We will
consider the effects of fluctuation-induced neutrino flavor
decoherence during two post-core-bounce epochs of supernova evolution:
(1) the shock reheating epoch at TPB (time post bounce) $< 1$ s; and
(2) the hot bubble $r$-process nucleosynthesis epoch at TPB $\approx$
3--16 s.

The neutrino heating of the stalled shock has been examined by Fuller
et al. [3] and Wilson and Mayle [4] including neutrino flavor-mixing
effects. Flavor mixing between a light $\nu_e$ and $\nu_{\mu}$ or
$\nu_{\tau}$ with a cosmologically significant mass in the range of
10--100 eV can increase the supernova explosion energy by up to 60\%.
The increase in the explosion energy is a result of the greater
temperature of $\nu_\mu$ and $\nu_\tau$ compared to that of
$\nu_e$. Because of their charged current interactions with nucleons,
$\nu_e$ remain in thermal equilibrium with the matter to lower
densities and temperatures than $\nu_\mu$ and $\nu_\tau$.  Thus,
although the neutrinos have approximately the same luminosities,
flavor transformation between $\nu_\mu$ or $\nu_\tau$ and $\nu_e$ on
their way to the stalled shock after thermal decoupling will increase
the average energy of $\nu_e$, resulting in more heat liberated behind
the shock than the case without neutrino flavor transformation.

This increase in the average energy of $\nu_e$ by the MSW
transformation also affects the possible $r$-process nucleosynthesis
of heavy elements in supernovae [5].  In the absence of flavor
transitions, $\bar\nu_e$ have a higher temperature than $\nu_e$. This
is due to the fact that the neutron-rich surface of the proto neutron
star presents a greater opacity to $\nu_e$ than $\bar\nu_e$.
Therefore, $\bar\nu_e$ decouple deeper in the proto neutron star than
$\nu_e$, and hence, are hotter. Due to their lack of charged current
interactions (at these energies) with nucleons, $\nu_\mu$ and
$\nu_\tau$ are even hotter than $\bar\nu_e$. As pointed out by Qian et
al. [5], an MSW transition can result in a proton production rate
which is greater than the neutron production rate at the proposed
$r$-process site.  By requiring neutron-rich conditions at the
$r$-process site, Qian et al.  were able to put limits on the vacuum
mass-squared difference and the vacuum mixing angle between $\nu_e$
and $\nu_\mu$ or $\nu_\tau$. Since $\nu_\mu$ and $\nu_\tau$ have
identical energy spectra, we will hereafter consider only MSW
transitions between $\nu_e$ and $\nu_\tau$ for convenience, although
all our results could equally well apply to MSW transitions between
$\nu_e$ and $\nu_\mu$.

A general, semi-quantitative approach to neutrino oscillations in
inhomogeneous matter was developed in Ref. [6].  Studies of matter
density fluctuations which are not random, but harmonic [7,8], or
occur as a jump-like change in the solar density [9], are available in
the literature.  Inhomogeneities in the velocity field of material can
mimic the effects of density inhomogeneities on neutrino flavor
oscillations [8]. Velocity inhomogeneity effects become important if
the characteristic velocity is near the speed of light [8]. Noisy
mixing of matter could also mimic a fluctuating matter density.

A priori, fluctuations in the matter density may be well approximated
by random noise (which averages to zero) added to the average value of
the density.  In Ref. [10], a differential equation for the averaged
survival probability was derived for the case in which the random
noise was taken to be a delta-correlated Gaussian distribution.  It
was also shown that if the correlation length of the matter density
fluctuations was small compared to the neutrino oscillation length at
resonance, one obtained the same result as for the case of a
delta-correlated Gaussian. Here, we consider the more realistic case
of colored noise [11]. We find that the random fluctuations have the
largest effect on the flavor transition when the correlation length is
on the order of the neutrino oscillation length at resonance, and that
the fluctuations, on average, have the effect of suppressing the
flavor transition. This result is in agreement with more qualitative
arguments regarding flavor transition in an inhomogeneous distribution
of matter [6].

Suppressing the degree of flavor transition has the effect of lowering
the average energy of $\nu_e$ compared with the average energy they
would have had for an MSW transition in the absence of the noise.  We
will show that neutrino flavor decoherence caused by random
fluctuations has little effect on the part of the neutrino mixing
parameter space $(\delta m^2,\
\sin^22\theta)$ previously excluded by considerations of MSW
neutrino flavor transformation in the hot bubble $r$-process region of
the supernova environment. However, we find that neutrino flavor
decoherence can have significant effects on the neutrino heating rate
and the electron fraction during the shock reheating epoch.

In Sec. II, we develop the equations for two-neutrino flavor
conversion in the presence of multiplicative colored noise.  We
present a differential equation for the averaged transition
probability, which is valid when the correlation length is small
compared to the width of the resonance region. Section III presents
our results for the effects of random fluctuations on neutrino heating
of the shock and $r$-process nucleosynthesis in the neutrino-heated
supernova ejecta. In Sec. IV, we discuss our results and present our
conclusions.

\section{NEUTRINO FLAVOR EVOLUTION IN THE PRESENCE OF COLORED NOISE}

In this section, we discuss how an initially pure neutrino flavor
state propagating through a stochastic field of density fluctuations
can evolve into a mixed ensemble of neutrino flavors. We call this
process neutrino flavor decoherence (or flavor depolarization). Here
we consider two-neutrino mixing and we assume that there is the usual
unitary transformation between flavor eigenstates (e.g.,
$|\nu_e\rangle$ and $|\nu_\tau\rangle$) and mass eigenstates
($|\nu_1\rangle$ and $|\nu_2\rangle$):

\begin{equation}
|\nu_e\rangle = \cos\theta|\nu_1\rangle+\sin\theta|\nu_2\rangle,
\end{equation}
\begin{equation}
|\nu_\tau\rangle=-\sin\theta|\nu_1\rangle+\cos\theta|\nu_2\rangle,
\end{equation}
where $\theta$ is the vacuum mixing angle. Complete flavor
decoherence would occur if a neutrino emitted from the neutrino sphere
in an initially pure flavor state, for example $|\nu_e\rangle$,
becomes a mixed state of 50\% $|\nu_e\rangle$ and 50\%
$|\nu_\tau\rangle$ (both at the original neutrino energy) after
propagating through a region where density fluctuations exist.

Our goal in this section is to find which fluctuation characteristics
(e.g., root-mean-square amplitude, power spectrum, etc.) would be
required to cause significant deviations in neutrino flavor evolution
from that predicted by conventional MSW studies with a smooth density
run in the post-core-bounce supernova environment. We begin with a
discussion of the general flavor evolution problem and the way in
which fluctuations can be characterized.

The density matrix of two-neutrino flavor evolution obeys the equation
\begin{equation}
i{d\over dr}\hat {\rho} = [\hat H,\hat {\rho}],
\end{equation}
where
\begin{equation}
\hat {\rho} \equiv \pmatrix{
a_e(r) \cr
a_{\tau}(r)\cr}\otimes
(a^*_e(r),\  a^*_{\tau}(r)),
\end{equation}
with $a_e$ and $a_{\tau}$ the probability
amplitudes for the neutrino
to be $\nu_e$ and $\nu_\tau$, respectively, and where the Hamiltonian is
given by \cite{r1}
\begin{equation}
\hat H = \left({{-\delta m^2}\over 4E} \cos 2\theta + {1\over \sqrt{2}} G_F
(N_e(r)
+ N^r_e(r))\right){\sigma_z} + \left({{\delta m^2}\over 4E}
\sin 2\theta \right) {\sigma_x}.
\end{equation}
In Eq. (5), $\delta m^2$ is the vacuum neutrino mass-squared
difference, $E$ is the neutrino energy, $N_e$ and $N^r_e$ are the
averaged and randomly fluctuating parts of the electron number
density, respectively, and $\sigma_x$ and $\sigma_z$ are the Pauli
matrices.

For colored noise, we take the ensemble averages of the randomly
fluctuating part of the electron density to be given by
\begin{equation}
\langle N^r_e(r)\rangle = 0,
\end{equation}
\begin{equation}
\langle N^r_e(r)N^r_e(r^{\prime}) \rangle = {\beta}^2  \ N_e(r)
\ N_e(r^{\prime})
 \ \exp(-|r-r^{\prime}|/\tau_c),
\end{equation}
with the averages of all odd products vanishing and all higher even
products given by all possible independent products of two-body
correlations (i.e., the fluctuations are Gaussian). For example, if we
define $f_{12 \cdots }=\langle N^r_e(r_1)N^r_e(r_2) \cdots \rangle$,
the average of a product of four would be $f_{1234}= f_{12}f_{34} +
f_{13}f_{24} + f_{14}f_{23}.$ In Eq. (7), $\beta$ is the ratio of the
root-mean-square fluctuation to the local density, and $\tau_c$ is the
correlation length.

Clearly, the detailed evolution of the position (time) dependent
amplitudes $a_e(r)$ and $a_\tau(r)$ through regions which include MSW
resonances (mass level crossings) is quite complicated in the presence
of fluctuations and so necessitates numerical treatment. However,
before we proceed to the description of the numerical calculation, it
is advantageous to define a few quantities which we shall later
employ. First, we introduce the resonance width as
\begin{equation}
\delta r=2H\tan2\theta,
\end{equation}
 where the effective weak charge
density scale height is $H\equiv|d\ln N_e(r)/dr|^{-1}$ [1]. We define
$\Delta\equiv\delta m^2/2E$, where $\delta m^2$ is the difference of
the squares of the vacuum neutrino mass eigenvalues. With this
notation, resonance (neutrino mass level crossing) occurs at the
position where $\Delta\cos2\theta=\sqrt{2}G_FN_e$ is satisfied.

We now proceed to a detailed description of how we follow numerically
the time evolution of the neutrino density matrix [i.e.,
Eq. (3)]. Provided one has access to a random number generator which
generates Gaussian deviates, Eq. (3) can be integrated by the method
of Ref. [11] with the probabilities calculated and averaged. However,
such a process can be quite time consuming. For the case in which the
correlation length is small compared with the width of the resonance
region, a ``ladder'' approximation can be made allowing one to obtain
a differential equation for the averaged probabilities.

 In order to obtain this approximation, one can transform $\hat
{\rho}$ to the interaction picture, express $\hat\rho_I$ as an
iterative expansion and explicitly perform the averaging to obtain
[10]
\begin{equation}
\langle\hat {\rho}_I(r)\rangle = {1\over 2}\left({1 + \sigma_z} \right)
+\sum_{n=1}^{\infty} (-1)^n \int dR(2n)
F(2n) \hat C (2n),
\end{equation}
where
\begin{equation}
\int dR(2n) \equiv \int_0^r dr_1 \int_0^{r_1} dr_2 \cdots \int_0^{r_{2n-1}}
dr_{2n},
\end{equation}
\begin{equation}
F(2n) \equiv f_{123 \cdots (2n-1) (2n)},
\end{equation}
\begin{equation}
\hat C (2n) \equiv [\hat M(r_1),[\hat M(r_2),[ \cdots, [\hat M(r_{2n}),
\sigma_z] \cdots],
\end{equation}
and $\hat M = (G_F/2\sqrt{2})\hat U_0^{\dagger} \sigma_z \hat U_0$,
with $\hat U_0$ being the propagator of the flavor evolution in the
absence of fluctuations. The averaged density matrix is then given by
$\langle\hat\rho\rangle=U_0\langle\hat\rho_I\rangle U_0^\dagger$.
{}From the physical fact that the fluctuations should not affect the
evolution far from resonance, one can restrict the integration limits
in Eq. (10) to the resonance region.  Since the matrix elements of
$\hat C(2n)$ are always $\leq 1$, we evaluate the integrals of the
terms in $F(2n)$ and compare the relative size of these terms.  When
$\tau_c \ll \delta r$, the leading term in the integral of $F(2n)$ is
of order $(\tau_c\delta r)^n$. This contribution comes from the term
in which the subscripts are in the order of the nested integrals
[i.e., ($f_{12}f_{34}
\cdots f_{(2n-1) (2n)})$]. Terms which are out of this order by $s$
interchanges are
of order $(\delta r)^{n-s} \tau_c^{n+s}$ when $\tau_c \ll \delta r$.
Note that because of the absolute value signs in Eq. (7),
$f_{ij} = f_{ji}$, so that $f_{13}f_{24}
\cdots f_{(2n-1) (2n)}$ is out of order by one interchange and
$f_{13}f_{45}f_{26}
\cdots f_{(2n-1) (2n)}$ is out of order by two interchanges.
Therefore, the integral of all terms in $F(2n)$
{\it not in} the order of the nested integrals will be of order
$\tau_c/\delta r$
or smaller than the integral of the term in $F(2n)$ which is {\it in} the order
of the nested integrals when $\tau_c \ll \delta r$. Therefore,
we approximate Eq. (9) by keeping only the largest term in
each $F(2n)$. One then
obtains an infinite series which is the iterative expansion of
\begin{eqnarray}
\langle\hat {\rho}_I(r)\rangle & = & {1\over 2}\left({1 + \sigma_z} \right)
- \int_0^r dr_1 \int_0^{r_1} dr_2\,
\beta^2 N_e(r_1) N_e(r_2) \exp[{-(r_1 - r_2)/\tau_c}] \nonumber\\
& \times & [\hat M(r_1), [\hat M(r_2),\langle\hat {\rho}_I(r_2)\rangle]],
\end{eqnarray}
where we have returned to the full integration limits since the
contribution from positions outside the resonance region is
small. Equation (13) could also have been obtained by making the
assumption $\langle N_e^r N_e^{r\prime}\hat {\rho}_I\rangle \sim
\langle N_e^r N_e^{r\prime}\rangle\langle\hat {\rho}_I\rangle$. Since
the maximal reduction in the transition probability occurs for cases
in which the correlation length is approximately the neutrino
oscillation length at resonance divided by $\pi$ (i.e., $\tau_c\sim
L_{\rm res}/\pi = 4E/\delta m^2 \sin 2\theta$), the above
approximation should be valid for adiabatic transitions for which the
oscillation length at resonance is much smaller than the width of the
resonance region.

As an example of the approximation, we calculate and present the
survival probability as a function of the correlation length for a
$\nu_\tau$ traveling through the supernova at TPB $\approx3$ s in
Fig. 1(a).  The rms value of the noise is taken to be 1\% of the local
electron number density, and we choose the following parameters:
$\delta m^2 = 10^2$ eV$^2$, $\sin^2 2\theta = 10^{-3}$, and $E = 33$
MeV. The correlation length of the random fluctuations varies from
$\tau_c =$ (0.25--24)$\tau_c^0$, where $\tau_c^0 = L_{\rm res}/\pi =
({{\delta m^2}}\sin 2\theta/4E)^{-1} = 832.6$ cm.  For these
parameters, the neutrino flavor evolution in the absence of noise is
highly adiabatic ($L_{\rm res}/ \delta r \sim 0.03$).  In Fig. 1, the
smooth curve is the approximation and the jagged line is the
simulation of Eq. (9) utilizing numerical averaging. The survival
probability in the absence of fluctuations is not shown in Fig. 1(a),
being $\sim 0$. The agreement between the approximation and the
simulation is quite good.  We also plot in Fig. 1(b) the $\nu_\tau$
survival probability as a function of the correlation length for the
same parameters as in Fig. 1(a), except with $\sin^2 2\theta =
10^{-5}$ and the corresponding $\tau_c^0 = 8326.0$ cm.  In Fig. 1(b),
the horizontal line is the survival probability in the absence of
fluctuations. One observes that at small values of the correlation
length [$\tau_c \sim$ (0--5)$\tau_c^0$], the survival probability from
the simulation is larger than that calculated from the approximation,
although they are not greatly different.

The reason the approximation works as well as it does is that for both
$\tau_c \gg L_{\rm res}$ and $\tau_c \ll L_{\rm res}$, the
approximation and the numerically-averaged simulation approach the
probability in the absence of fluctuations. As discussed in Ref. [10],
the parameter $\gamma = 1/2 G_F^2 \langle(N_e^r)^2\rangle \tau_c
\delta r$ governs the size of the effect of the fluctuations when
$\tau_c \ll L_{\rm res}$. When $\gamma$ is large, the transition
probability is heavily suppressed, where as for $\gamma \ll 1$, the
fluctuations have little effect. Therefore, a small value for $\tau_c$
will result in only a small change in the probability. As $\tau_c$
goes to infinity,
\begin{eqnarray}
\lim_{\tau_c\to\infty}F(2n) & = & (2n-1)!! \beta^{2n}
\prod_{i=1}^{2n}
N_e(r_i) \nonumber \\
& = & {1\over{\sqrt{2\pi \beta^2}}} \int_{-\infty}^{\infty} dx
\exp[{-x^2/(2\beta^2)}] x^{2n} \prod_{i=1}^{2n} N_e(r_i),
\end{eqnarray}
and Eq. (9) can be summed and transformed back from the
interaction picture to give
\begin{equation}
\lim_{\tau_c\to\infty}\langle\hat \rho(r)\rangle =
{1\over{\sqrt{2\pi \beta^2}}} \int_{-\infty}^{\infty} dx
\exp[{-x^2/(2\beta^2)}]
\hat \rho(r,x),
\end{equation}
where $\hat \rho(r,x)$ is the density matrix calculated using Eq. (3)
for the electron density $(1 + x)N_e(r)$. Equation (15) provides a
simple physical picture for the averaged density matrix in the limit
of very large correlation lengths: one simply calculates the density
matrix for the electron density with a ``frozen'' fluctuation,
$(1+x)N_e(r)$, then averages over all such fluctuations with a
Gaussian weight. Large correlation lengths imply that fluctuations at
many different locations are coupled. Indeed, an averaging such as
given in Eq. (15) is common in other physical situations when many
coupled channels are present. For example, in multidimensional
dissipative quantum tunnelling, barrier transmission probability is
given by a similar formula when the number of channels gets very large
[12]. Note that to first order in $x$, the change
$N_e(r)\to(1+x)N_e(r)$ in the functional form of the density will
affect the survival probability via a change in the slope of the
density at resonance. For slowly changing slopes and small $\beta$,
the change in the survival probability should be proportional to
$\beta^2$, and therefore be quite small.

\section{EFFECTS ON THE $r$-PROCESS AND \hfill \break SHOCK REHEATING}

As demonstrated by Fig. 1(b), even a 1\% fluctuation in the matter
density (a value which is probably unrealistically large for the
rather quiescent TPB $>3$ s epoch) does little in reducing the
transition probability (or increasing the survival probability) for
non-adiabatic evolution. As shown in Ref. [5], the neutrino-heated
supernova ejecta must have a neutron excess ($Y_e<0.5$) in order for
any $r$-process nucleosynthesis to be produced. In Ref. [5], it was
shown that the electron fraction $Y_e$ is approximately given by
\begin{equation}
Y_e \approx {1 \over {1 + \lambda_{\bar{\nu}_e p}/
\lambda_{\nu_e n}}} \approx {1 \over {1 + \langle{E}_{\bar{\nu}_e}\rangle/
\langle{E}_{\nu_e}\rangle}},
\end{equation}
where $\lambda_{\bar{\nu}_e p}$ and $\lambda_{\nu_e n}$ are the
reaction rates for
\begin{eqnarray}
\bar{\nu}_e + p & \rightarrow & e^+ + n,\\
\nu_e + n & \rightarrow & e^- + p,
\end{eqnarray}
respectively, and $\langle E_{\nu_e}\rangle$ and
$\langle{E}_{\bar\nu_e}\rangle$ are the average energy for $\nu_e$ and
$\bar\nu_e$, respectively.  Since at this epoch the average neutrino
energies are $\langle E_{\nu_e}\rangle\approx11$ MeV, $\langle
E_{\bar\nu_e}\rangle\approx16$ MeV, and $\langle
E_{\nu_\tau}\rangle\approx25$ MeV, a substantial conversion of
$\nu_\tau$ into $\nu_e$ can decrease the ratio
$\langle{E}_{\bar{\nu}_e}\rangle/ \langle{E}_{\nu_e}\rangle$ below one
and thereby make conditions impossible for the $r$-process. The
average $\nu_e$ energy after an MSW transition is approximately
$P(\nu_e \rightarrow \nu_e) \langle{E}_{\nu_e}\rangle + P(\nu_{\tau}
\rightarrow \nu_e)\langle{E}_{{\nu}_\tau}\rangle$. To obtain
neutron-rich conditions, one must have $P(\nu_e \rightarrow \nu_e) >
64\%$. It only requires about 30\% efficiency in flavor conversion to
drive the neutrino-heated supernova ejecta too proton rich for
$r$-process nucleosynthesis. Consider simple estimates for $Y_e$ using
Eq. (16) for four cases:

(1) No flavor mixing. In this case, we have $Y_e\approx1/(1+\langle
E_{\bar\nu_e}\rangle/ \langle
E_{\nu_e}\rangle)\approx1/(1+16/11)\approx0.41$, so the material is
neutron rich, in good agreement with detailed supernova model
calculations [5].

(2) Full flavor conversion. In this case, we have $\langle
E_{\nu_e}\rangle\rightleftharpoons\langle E_{\nu_\tau}\rangle$, so
that $Y_e\approx1/(1+\langle E_{\bar\nu_e}\rangle/ \langle
E_{\nu_\tau}\rangle)\approx1/(1+16/25)\approx0.61,$ and the material
is very proton rich.

(3) Complete neutrino flavor depolarization. In this case, there is
50\% flavor conversion, so $\langle E_{\nu_e}\rangle\to(\langle
E_{\nu_e}\rangle+\langle E_{\nu_\tau}\rangle)/2\approx18$ MeV, which
implies that $Y_e\approx1/(1+\langle E_{\bar\nu_e}\rangle/\langle
E_{\nu_e}^\prime\rangle)\approx0.53,$ too proton rich for $r$-process
nucleosynthesis.

In fact, we must have $Y_e<0.5$ to get {\it any} $r$-process
nucleosynthesis. This is a conservative limit, since a {\it good}
$r$-process requires $Y_e\leq0.45$. Now consider a fourth case,

(4) Partial flavor depolarization. For 35\% conversion of $\nu_\tau$
into $\nu_e$, we have $Y_e\approx 0.5$. For 30\% conversion of
$\nu_\tau$ into $\nu_e$, we have $Y_e\approx 0.49,$ realistically too
large to give acceptable $r$-process nucleosynthesis. In fact, to get
$Y_e<0.45$ we need to demand that there had been less than 15\% flavor
conversion.

Consider some condition along the $Y_e=0.5$ line in Fig. 2 of
Ref. [5]. For example, consider $E_\nu=25$ MeV and $\delta m^2\approx
900$ eV$^2$, for which the density scale height at resonance is
$H\approx0.5$ km. This value of $\delta m^2$ corresponds to
$\sin^22\theta\approx4\times10^{-6}$ on the $Y_e=0.5$ line. In this
case, the conversion probability is
$P({\nu_\tau\to\nu_e})\approx1-\exp\{-0.04\left({\delta m^2/ {\rm
eV}^2}\right)\left({{\rm MeV}/ E_\nu}\right)\left({H/{\rm
cm}}\right)\sin^22\theta\}\approx 25\%$. The neutrino energies around
25 MeV are the most important in terms of leverage on $Y_e$. Note that
fluctuation-induced depolarization at a level of 50\% could
conceivably produce {\it more} conversion than MSW transformation at
$E_\nu=25$ MeV over a considerable part of the region to the {\it
right} of the $Y_e=0.5$ line in Fig. 2 of Ref. [5]. As outlined above,
greater than 30\% flavor conversion will always drive the material too
proton rich for $r$-process.

One can conclude that the random fluctuations will have a quite minor
effect on the neutrino mixing parameters constrained by $r$-process
nucleosynthesis because the absolute increase of the survival
probability rapidly diminishes with increasing survival
probability. To illustrate this, we plot the survival probability as a
function of energy for $\nu_\tau$ with fluctuations of 1\% and 0.5\%
of the local matter density in Figs. 2(a) and 2(b). We choose the
parameters of Fig. 1(b) so that the evolution is nonadiabatic, and the
neutrino energy is chosen to maximize the differential capture rate
[see Eq. (23)]. In Fig. 2, the jagged lines are the solution using
Eq. (9), the solid line is the approximate solution, and the dashed
line is the survival probability in the absence of fluctuations. One
observes that the approximate solution does fairly well in reproducing
the simulation. There is a small increase of about $0.08$ in the
survival probability in Fig. 2(a) where the fluctuations are 1\% of
the local density, but only a very small increase is obtained for
0.5\% noise in Fig. 2(b). If one calculates $Y_e$ for these parameters
in the absence of fluctuations, one obtains $Y_e = 0.5$ indicating
that this point lies on the boundary of the excluded region of the MSW
parameter space [5]. The inclusion of noise at the 1\% level would
decrease $Y_e$ by about 1\%. Random fluctuations of 0.5\% as in
Fig. 2(b) would give a decrease of about 0.4\%.  For noise at the
0.1\% level, there will essentially be no change in the excluded
region.

Note that if the neutrino flavor transition is adiabatic, the
fluctuations can increase the survival probability from approximately
zero to one half. The shock reheating epoch occurs at approximately
TPB $\sim0.15$ s, and the relevant scale heights are larger than those
at TPB $>3$ s, which implies a larger resonance width for comparable
values of $\delta m^2$ and $\sin^22\theta$. Hence, $\gamma$ may remain
$\sim 1$ while $\langle(N_e^r)^2\rangle$ is reduced allowing one to
obtain a sizable effect for adiabatic transitions as shown in
Fig. 1(a) for TPB $\approx3$ s with a smaller rms value of the noise.
In Fig. 3 we show the survival probability as a function of neutrino
energy for the values $\delta m^2 = 10^3$ eV$^2$, $\sin^2 2\theta =
10^{-6}$, $\tau_c = 2.762 \times 10^3$ cm. The rms fluctuations are
0.05\% and 0.1\% of the local density in Figs. 3(a) and 3(b),
respectively. In both cases there is a large region where the survival
probability is increased from the value of essentially zero in the
absence of fluctuations. The approximate solution seems to be better
in Fig. 3(b) where the rms value of the fluctuations is larger, but is
reasonably close to the numerical simulation in both cases.  For the
case of an adiabatic transition between the more energetic $\nu_\tau$
and less energetic $\nu_e$, the heating rate can be increased by
(30--60)\% [3]. Therefore, the reduction of the transition probability
by random fluctuation effects for an a priori reasonable rms value of
the fluctuations [(0.05--0.1)\%] could be important.

We can estimate the decrease in the heating rate due to the
fluctuations over that from the case of an adiabatic MSW transition as
follows. In the absence of neutrino flavor conversion, the heating
rate per nucleon is approximately given by [13]
\begin{equation}
\dot {\epsilon}_{\nu N} \approx {{L_{\nu}}\over {4 \pi r^2}}
{{\int_0^{\infty} E^3_{\nu} \sigma_{\nu N}
dE_{\nu}/(\exp({E_{\nu}/T_{\nu}}) + 1)} \over
{\int_0^{\infty}E^3_{\nu} dE_{\nu}/(\exp({E_{\nu}/T_{\nu}}) + 1)}} =
\int_0^{\infty} E_{\nu} \sigma_{\nu N} {{d\phi(E_{\nu},T_{\nu})}\over
dE_{\nu}} dE_{\nu},
\end{equation}
where
\begin{equation}
\sigma_{\nu N} \approx 9.6 \times 10^{-44} \left( {E_e \over {\rm MeV}}
\right)^2 {\rm cm^2},
\end{equation}
is the absorption cross section, $L_{\nu}$ is the neutrino luminosity
(assumed equal for all species), $T_{\nu}$ is the neutrino
temperature, $d\phi$ is the differential neutrino flux with respect to
neutrino energy, and $E_e$ is the energy of the produced electron or
positron in Eqs. (17) and (18).

The total heating rate combining contributions from both $\nu_e$ and
$\bar\nu_e$ is
\begin{equation}
\dot {\epsilon}_{\rm tot} = Y_n \dot {\epsilon}_{\nu_e n} + Y_p \dot
{\epsilon}_{\bar{\nu}_e p},
\end{equation}
where $Y_n$ and $Y_p$ are the number fractions of free neutrons and
protons, respectively, with $Y_n + Y_p \approx 1$. Of course, $Y_n$
and $Y_p$ will be set by the same weak interactions responsible for
heating the shock. In the region where neutrino heating dominates, we
can assume
\begin{equation}
{Y_n \over Y_p} \approx {{\lambda_{\bar{\nu}_e p}} \over
{\lambda_{\nu_e n}}},
\end{equation}
where $\lambda_{\nu N}$ is the neutrino capture rate on free nucleons and
is given by
\begin{equation}
\lambda_{\nu N} \approx
\int_0^{\infty} \sigma_{\nu N} {{d\phi(E_{\nu},T_{\nu})}\over dE_{\nu}}
dE_{\nu}.
\end{equation}

An MSW transition between $\nu_e$ and $\nu_\tau$ will increase the
heating rate by a factor of
\begin{equation}
{{\dot {\epsilon}_{tot}^{\prime}}\over {\dot {\epsilon}_{tot}}} = {{
Y_n^{\prime} \dot {\epsilon}_{\nu_e n}^{\prime} + Y_p^{\prime} \dot
{\epsilon}_{\bar{\nu}_e p}} \over {Y_n \dot {\epsilon}_{\nu_e n} +
Y_p \dot {\epsilon}_{\bar{\nu}_e p}}},
\end{equation}
where $\dot {\epsilon}_{\nu_e n}^{\prime}$ is given by
\begin{eqnarray}
\dot {\epsilon}_{\nu_e n}^{\prime} & = &
\int_0^{\infty} P(E_{\nu},\tau_c,\beta ) E_{\nu} \sigma_{\nu N}
{{d\phi(E_{\nu},T_{\nu_e})}\over dE_{\nu}} dE_{\nu} \nonumber \\
& + & \int_0^{\infty}[1 - P(E_{\nu},\tau_c,\beta )] E_{\nu} \sigma_{\nu N}
{{d\phi(E_{\nu},T_{\nu_{\tau}})}\over dE_{\nu}} dE_{\nu}.
\end{eqnarray}
We take $Y_n^{\prime}$ and $Y_p^{\prime}$ to be given by Eq. (22) with
\begin{eqnarray}
\lambda_{\nu_e n}^{\prime} & = &
\int_0^{\infty} P(E_{\nu},\tau_c,\beta ) \sigma_{\nu N}
{{d\phi(E_{\nu},T_{\nu_e})}\over {dE_{\nu}}} dE_{\nu} \nonumber \\
& + &
\int_0^{\infty}[1 - P(E_{\nu},\tau_c,\beta )] \sigma_{\nu N}
{{d\phi(E_{\nu},T_{\nu_{\tau}})}\over {dE_{\nu}}} dE_{\nu}.
\end{eqnarray}
In the above equations, $P(E_{\nu},\tau_c,\beta)$ is the survival
probability of $\nu_\tau$ (or $\nu_e$).

We have calculated Eq. (24) for the cases of random fluctuations with
an rms value of $0.05 \%$ and $0.1 \%$ of the local matter density,
and for the case without fluctuations. We present the results in Table
I. For {\it each} set of values for $(\delta m^2,\ \sin^2 2\theta)$,
we have taken $\tau_c = ({{\delta m^2 \sin 2\theta} / {4 E_{\rm
peak}}} )^{-1}$, where $E_{\rm peak} = 5 T_{\nu_{\tau}}$, the energy
for which the integrand is approximately maximized. We have taken the
temperatures to be $T_{\nu_e} = T_{\bar{\nu}_e} \approx 5$ MeV and
$T_{\nu_{\tau}} \approx 7$ MeV.  The values in Table I were calculated
using the approximation in Eq. (13) rather than the ``exact''
numerical simulation, since we are interested in estimating the size
of the difference in neutrino flavor conversion efficiency for the
cases of smooth and noisy density distributions, not on a particular
model of the fluctuations which may or may not obey Eqs. (6) and (7).
{}From Table I, one sees that the noise can reduce the heating rate by
up to 45\% from the adiabatic MSW case.

\section{DISCUSSION AND CONCLUSIONS}

To our knowledge, there is no consensus on the size of possible matter
density fluctuations in post-core-bounce supernovae. Although we have
taken the rms size of the fluctuations to be a constant fraction of
the local matter density, it may very well increase or decrease with
decreasing density, or the size could depend on the distance from the
shock.  Similarly, the correlation lengths we have used were chosen to
give the maximal reduction in the MSW transition probability and were
independent of the density and the distance from the shock. If, for
instance, the correlation length increased with decreasing density,
the effect of the noise in reducing the MSW transition probability
could be enhanced since for fixed $\delta m^2$ and $\sin^2 2\theta$
the correlation length of maximal effect ($\tau_c \sim L_{\rm res}$)
varies linearly with the neutrino energy. Our intention in this paper
is to establish the maximal effect density fluctuations could have if
they can be well approximated by random noise added to the average
density.

For an rms fluctuation value of 1\%, the addition of the noise will
have a slight effect on the $r$-process nucleosynthesis as compared
with the MSW effect in the absence of noise.  This can be traced to
the fact that demanding neutron-rich conditions at the site of the
$r$-process eliminates all but the nonadiabatic region of the neutrino
mixing parameter space. The effect of the noise is large in the
adiabatic region only and becomes increasingly negligible as the
neutrino flavor transition becomes nonadiabatic. A 1\% fluctuation
will decrease $Y_e$ from $0.5$ to $\sim 0.495$ for points in the MSW
parameter space on the boundary of the excluded region [5].  However,
we consider this rms fluctuation value to be actually too large for
the relatively quiesent TPB $>3$ s epoch.  For an order of magnitude
smaller fluctuation which may be more reasonable, the effect of the
noise is totally ignorable. In other words, noise at the 0.1\% level
will not alter the region of the neutrino mixing parameter space
excluded in Ref. [5] by the MSW effect.

Noise with an rms amplitude of 0.05\% of the averaged local matter
density can lead to significant neutrino flavor decoherence during the
shock reheating epoch at TPB $\sim0.15$ s. In turn, this would lead to
a dimunition of the adiabatic MSW-induced increase in the supernova
explosion energy from $\sim$ (30--60)\% to $\sim 20\%$. The physics of
supernova explosion process is quite complicated and depends on many
factors.  In our opinion, if the heating of the stalled shock by an
MSW transition is determined to be a necessary component for a
successful explosion, then a more detailed analysis of the effects of
random matter density fluctuations on the MSW transition would be
warranted.

\vskip .2in
We wish to thank Ray Sawyer for bringing this problem to our attention
and providing many valuable insights. We also want to thank Wick
Haxton and Lincoln Wolfenstein for very useful discussions. This work
was supported by the Department of Energy under Grant
No. DE-FG06-90ER40561 at the Institute for Nuclear Theory, by NSF
Grant No. PHY-9503384 at UCSD, by NSF Grant No. PHY-9314131 and by the
Wisconsin Alumni Research Foundation at UW.
\vfill
\eject

\vfill
\eject

\setcounter{page}{18}
\noindent
{\bf Table I.} Ratios of the heating rates in the presence of MSW transitions
with no fluctuations ($\beta=0$), 0.05\% fluctuations ($\beta = 5
\times 10^{-4}$), and 0.1\% fluctuations ($\beta = 10^{-3}$) to the heating
rate in the absence of flavor transitions. The neutrino temperatures are
$T_{\nu_e} =
T_{\bar{\nu}_e} = 5$ MeV and $T_{\nu_{\tau}} =  7$
MeV.
\vskip 0.2in
\begin{tabular}{ccccc}  \hline\hline
& & &${\dot {\epsilon}_{\rm tot}^{\prime}} / {\dot {\epsilon}_{\rm tot}}$& \\
$\delta m^2 ({\rm eV^2})$ & $\sin^2 2\theta$
 & $\beta = 0$
& $\beta = 5 \times 10^{-4}$ & $\beta = 10^{-3}$ \\ \hline
 & & & & \\
$5 \times 10^2$ & $2 \times 10^{-8}$ & $1.05$ & $1.05$ & $1.05$ \\
 & & & & \\
$5 \times 10^2$ & $ 10^{-7} $ & $1.16$ & $1.14$ & $1.12$ \\
 & & & & \\
$5 \times 10^2$ & $ 10^{-6} $ & $1.31$ & $1.23$ & $1.18$ \\
 & & & & \\
$10^3$ & $2 \times 10^{-8}$ & $1.07$ & $1.07$ & $1.06$ \\
 & & & & \\
$10^3$ & $ 10^{-7} $ & $1.23$ & $1.19$ & $1.17$ \\
 & & & & \\
$10^3$ & $ 10^{-6} $ & $1.35$ & $1.23$ & $1.19$ \\
 & & & & \\
$8 \times 10^3$ & $2 \times 10^{-8}$ & $1.12$ & $1.11$ & $1.11$ \\
 & & & & \\
$8 \times 10^3$ & $ 10^{-7} $ & $1.31$ & $1.19$ & $1.19$ \\
 & & & & \\
$8 \times 10^3$ & $ 10^{-6} $ & $1.35$ & $1.19$ & $1.19$ \\ \hline\hline
\end{tabular}
\vfill
\eject

\vskip .2in
\centerline{\bf Figure Captions}
\vskip .2in
\noindent
{\bf Figure 1.} Survival probability $P(\nu_{\tau} \rightarrow
\nu_{\tau})$ is given as a function of the correlation length,
$\tau_c$ for the numerical simulation by the method of Ref. [11] and
by the approximation in Eq. (9). The parameters chosen were $\delta
m^2 = 10^2$ eV$^2$, $\sin^2 2\theta = 10^{-3}$, $E = 33$ MeV, and
$\tau_c^0 = L_{\rm res}/\pi = ({{\delta m^2}}\sin 2\theta/4E)^{-1} =
832.6$ cm for (a). The values $\delta m^2 = 10^2$ eV$^2$, $\sin^2
2\theta = 10^{-5}$, $E = 33$ MeV, and $\tau_c^0= 8326$ cm were chosen
for (b).  The rms value of the fluctuations was 1\% of the local
density in both (a) and (b). The matter density profile at TPB
$\approx3$ s was used [5] for both (a) and (b). The horizontal line in
(b) is the survival probability in the absence of fluctuations.
\vskip .2in
\noindent
{\bf Figure 2.} Survival probability $P(\nu_{\tau} \rightarrow
\nu_{\tau})$ is given as a function of energy for the numerical
simulation by the method of Ref. [11] and by the approximation in
Eq. (9). The rms value of the fluctuations is taken to be 1\% in (a)
and 0.5\% in (b). In both (a) and (b) the survival probability in the
absence of the fluctuations is shown by the dashed line.  The
parameters chosen were $\delta m^2 = 10^2$ eV$^2$, $\sin^2 2\theta =
10^{-5}$, $\tau_c = 8326$ cm.  The matter density profile at TPB
$\approx3$ s was used [5].
\vskip .2in
\noindent
{\bf Figure 3.} Survival probability $P(\nu_{\tau} \rightarrow
\nu_{\tau})$ is given as a function of energy for the numerical
simulation by the method of Ref. [11] and by the approximation in
Eq. (9). The rms value of the fluctuations is 0.05\% in (a) and 0.1\%
in (b). In both (a) and (b) the survival probability in the absence of
the fluctuations is essentially zero.  The parameters chosen were
$\Delta m^2 = 10^3$ eV$^2$, $\sin^2 2\theta = 10^{-6}$, $\tau_c =
2792.4$ cm.  The matter density profile at TPB $\approx0.15$ s was
used [4] and the density profile was cut off at a radius corresponding
to the position of the stalled shock.
\vfill
\eject

\epsfysize=6.5in \epsfbox{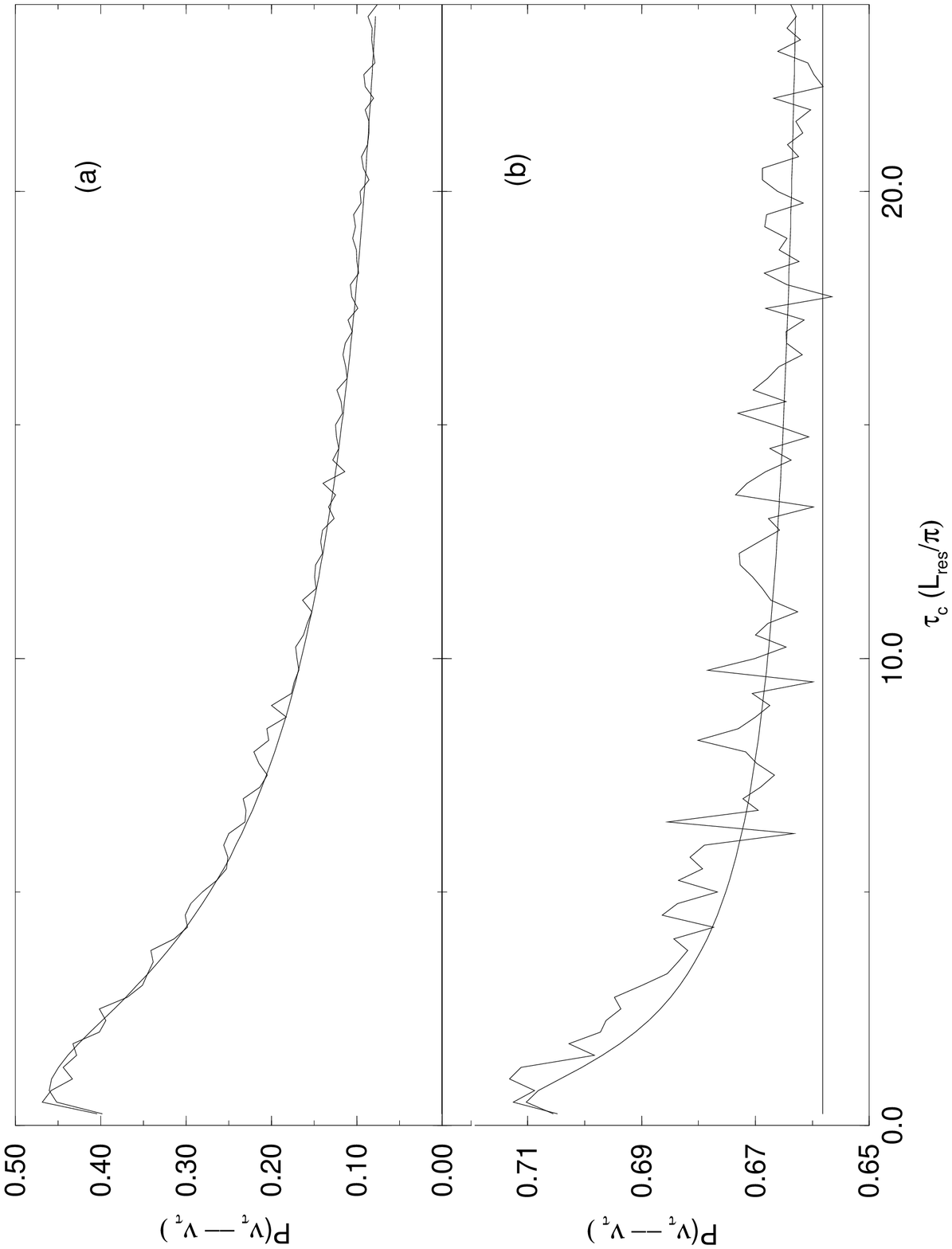}
\epsfxsize=6in \epsfbox{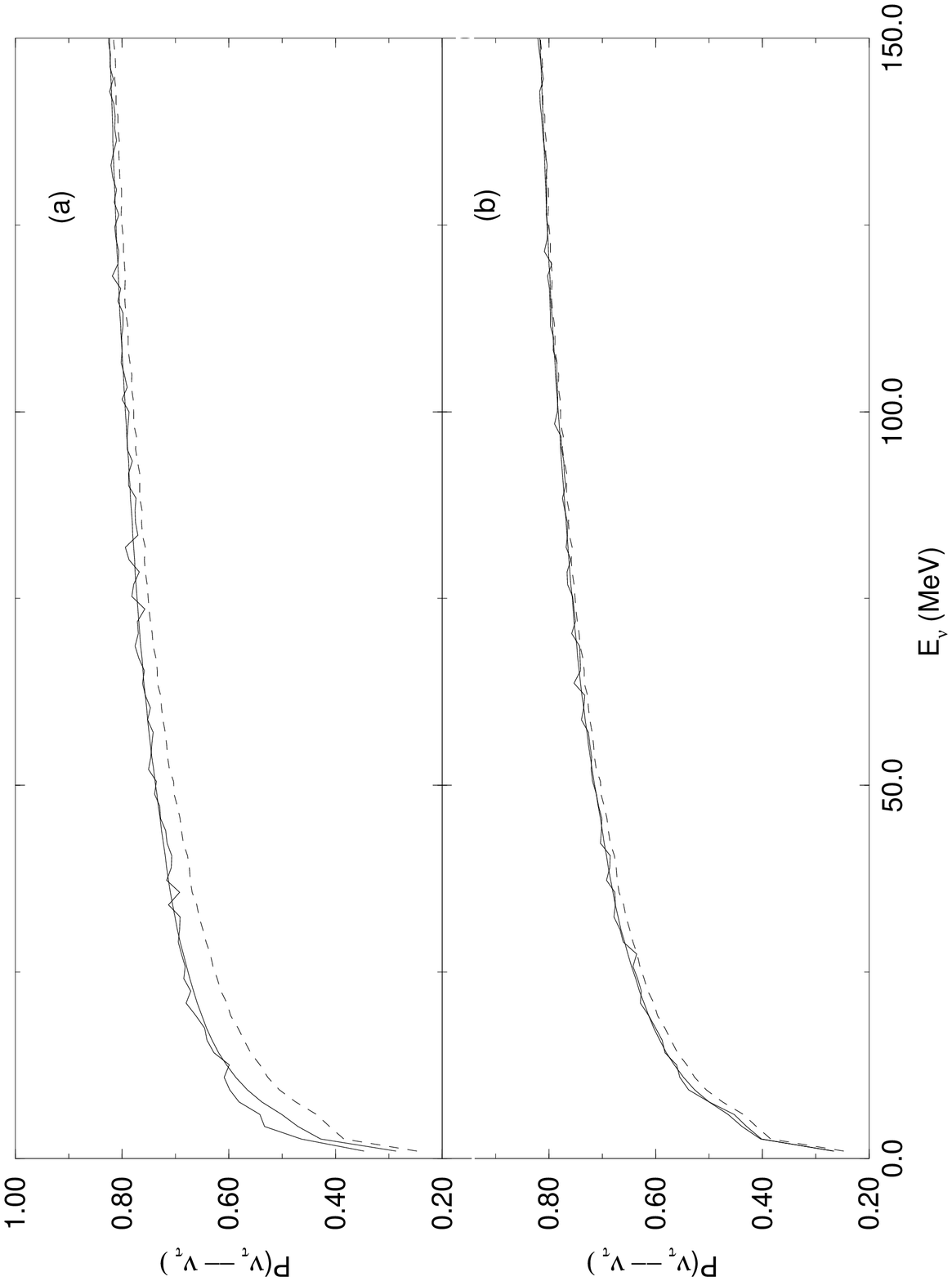}
\epsfysize=7in \epsfbox{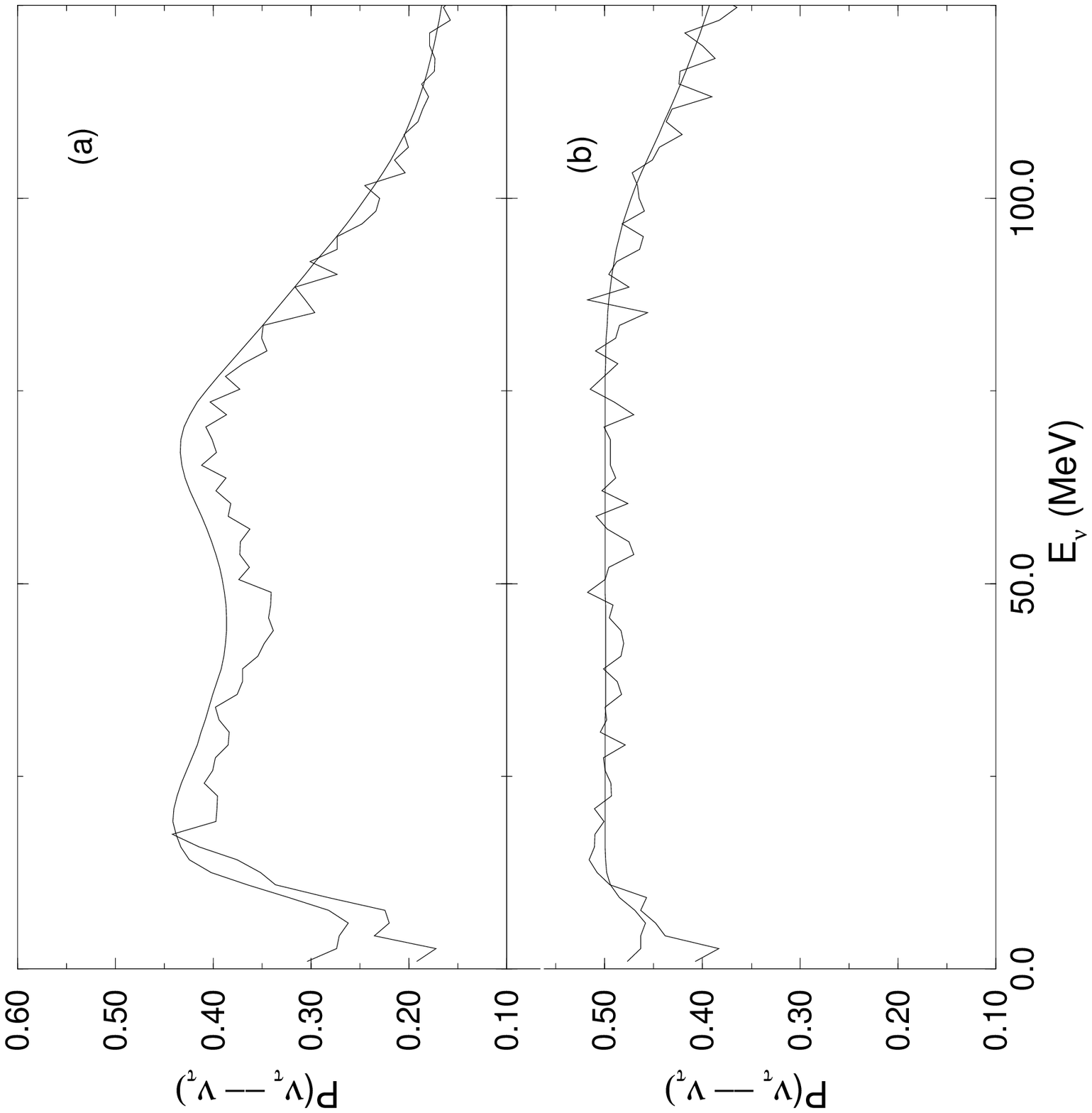}

\end{document}